\documentclass[aps,prb,twocolumn,showpacs]{revtex4}
\usepackage{graphicx}
\usepackage{amsmath,amssymb,amsfonts}
\usepackage{natbib}

\newcommand{\bd}{\begin{displaymath}}
\newcommand{\ed}{\end{displaymath}}
\newcommand{\be}{\begin{equation}}
\newcommand{\ee}{\end{equation}}
\newcommand{\bs}{\begin{subequations}}
\newcommand{\es}{\end{subequations}}
\newcommand{\ba}{\begin{eqnarray}}
\newcommand{\ea}{\end{eqnarray}}

\begin{document}

\title{Stochastic theory of lineshape broadening in quasielastic He
atom scattering\\ with interacting adsorbates}

\author{R. Mart\'{\i}nez-Casado}
\email{ruth@imaff.cfmac.csic.es}

\affiliation{Lehrstuhl f\"ur Physikalische Chemie I,
Ruhr-Universit\"at Bochum, D-44801 Bochum, Germany}

\affiliation{Instituto de F\'{\i}sica Fundamental,
Consejo Superior de Investigaciones Cient\'{\i}ficas,
Serrano 123, 28006 - Madrid, Spain}

\author{J.L. Vega}
\email{jlvega@imaff.cfmac.csic.es}

\author{A.S. Sanz}
\email{asanz@imaff.cfmac.csic.es}

\author{S. Miret-Art\'es}
\email{s.miret@imaff.cfmac.csic.es}

\affiliation{Instituto de F\'{\i}sica Fundamental,
Consejo Superior de Investigaciones Cient\'{\i}ficas,
Serrano 123, 28006 - Madrid, Spain}

\date{\today}

\begin{abstract}
The activated surface diffusion of interacting adsorbates is described
in terms of the so-called interacting single adsorbate approximation,
which is applied to the diffusion of Na atoms on Cu(001) for coverages
up to 20\% in quasielastic He atom scattering experiments.
This approximation essentially consists of solving the standard
Langevin equation with two noise sources and frictions: a Gaussian white noise
accounting for the friction with the substrate, and a white shot noise
characterized by a collisional friction simulating the
adsorbate-adsorbate collisions.
The broadenings undergone by the quasielastic peak are found to be
in very good agreement with the experimental data reported at two
surface temperatures 200 and 300~K.
\end{abstract}





\maketitle


\section{Introduction}
\label{sec1}

The diffusion of atoms, molecules, or small clusters on metal surfaces
plays an important role in many technological and industrial
applications.
Molecular beam epitaxy, heterogeneous catalysis, or the fabrication
of nanostructures are examples strongly affected by the kinetics of
diffusion.
Among the different experimental techniques utilized to analyze the
diffusion and adsorbate dynamics, quasielastic helium-atom scattering
(QHAS) is a gentle and inert technique commonly applied to fast
diffusion.\cite{salva1}
QHAS has been applied to different metal/metal and adsorbate/metal
systems, the diffusion of Na adatoms (at different coverages) on
Cu(001) being one of the most extensively systems studied so
far.\cite{toennies1,toennies2,josele1,josele2,josele3,ruth1,ruth2,ruth3,
allison,ruth4,ruth5}
At low coverages, this system has been theoretically described within
the so-called single adsorbate approximation and analyzed in
terms of the {\it motional narrowing effect},\cite{josele1,ruth1} and
also within the Kramers' turnover theory\cite{josele2} and the
dephasing theory.\cite{josele3}
In particular, the motional narrowing effect governs the broadening of
the lineshapes as a function of the friction, the parallel momentum
transfer, the lattice structure, and the jump dynamics.
Generalized expressions for the lineshape have been recently given
under these different conditions.\cite{ruth1,ruth2}

When dealing with higher coverages, absorbate-adsorbate interactions
can no longer be neglected.
Then, pairwise interaction potentials are usually introduced into
two-dimensional Langevin molecular dynamics (LMD) simulations,
\cite{toennies2} where the number
of coupled equations to be solved is usually very high (it increases
as $2N$, where $N$ is the number of adparticles considered, typically
of the order of 400-500).
On the one hand, first, this translates into a high computational cost.
Second, the numerical results issued from LMD simulations are difficult
to interpret since analytical treatments are not easily implementable.
On the other hand, from a physical viewpoint, the broadening (measured
as the full width at half maximum) of both the quasielastic ({\it Q})
peak ruling the diffusion process and the {\it T}-mode peaks related
to the low frequency motions of the adsorbate ({\it frustrated
translational modes}), is not well reproduced when compared with
the experimental data.\cite{toennies2}
Recently, it has been shown\cite{allison} that a good agreement with
the experiment can be achieved using simple models where the adsorbate
is allowed to also move perpendicular to the surface.

Aimed to provide a theoretical and numerical alternative to the
standard procedure at intermediate coverages, we have recently
proposed the so-called interacting single adsorbate (ISA)
approximation.\cite{ruth3,ruth4,ruth5}
Within this approach, diffusion is described by using {\it only one}
standard Langevin equation, which is characterized by the following
three contributions: (1) the deterministic, adiabatic potential $V$,
which models the adsorbate-substrate interaction at $T=0$; (2) a
Gaussian white noise, $R_G(t)$, accounting for the lattice vibrational
effects that the surface temperature induces on the adsorbate; and (3)
a white shot noise, which stands for the adsorbate-adsorbate collisions
and replaces the pairwise (dipole-dipole) interaction potential
generally considered.
In this way, a typical LMD simulation involving $N$ adsorbates is
substituted by the dynamics of a single adsorbate, with the action of
the remaining $N$-1 adsorbates being described by the random force
associated with the white shot noise.
This noise describes the adsorbate-adsorbate interaction by a series of
random pulses within a Markovian regime (i.e., pulses of relatively
short duration in comparison with the system relaxation).
This interaction is, therefore, described in terms of a
{\it collisional friction} which depends on the surface coverage.
With this simple stochastic model, a better agreement with the
experimental data for coverages up to around 20\%, approximately,
is obtained.
Although further investigation at microscopic level and calculations
from first principles are needed, this simple stochastic model at
moderate coverages is also able to provide a complementary view of
diffusion and low frequency vibrational motions, described by the
peaks at or around zero energy transfers (very long time dynamical
processes), respectively.
This could be understood because any trace of the true interaction
potential seems to be wiped out due to the relatively large number
of collisions taking place at very long times.

The organization of this paper is as follows.
In Sec.~\ref{sec2}, we introduce the stochastic model used in the ISA
approximation as well as its connection with the observable magnitudes
in QHAS experiments.
In Sec.~\ref{sec3}, the working model considered for the diffusion of
Na on Cu(001) is described, and numerical results are presented and
discussed.
Final remarks and conclusions are summarized in Sec.~\ref{sec4}.


\section{Interacting single adsorbate approximation}
\label{sec2}

In analogy to scattering of slow neutrons by crystals and
liquids,\cite{Mcquarrie,vanHove} the observable magnitude in QHAS
experiments is the {\it differential reflection coefficient},
\be
 \frac{d^2  {\mathcal R} (\Delta {\bf K}, \omega)}{d\Omega d\omega}
  = n_d {\mathcal F} S(\Delta {\bf K}, \omega) .
 \label{eq:DRP}
\ee
This coefficient gives the probability that the He atoms scattered from
the adsorbates on the surface reach a certain solid angle $\Omega$
with an energy exchange $\hbar\omega =E_f - E_i$ and wave vector
transfer parallel to the surface $\Delta {\bf K} = {\bf K}_f -
{\bf K}_i$.
In Eq.~(\ref{eq:DRP}), $n_d$ is the concentration of adparticles;
${\mathcal F}$ is the {\it atomic form factor}, which depends on the
interaction potential between the probe atoms in the beam and the
adparticles on the surface; and $S(\Delta {\bf K},\omega)$ is the
{\it dynamic structure factor} or {\it scattering law} which gives the
{\it Q} and {\it T} peaks and it provides a complete information about
the dynamics and structure of the adsorbates through particle
distribution functions.
Experimental information about long distance correlations
is obtained from the scattering law when considering small values of
$\Delta {\bf K}$, while information on long time correlations is
provided at small energy transfers, $\hbar \omega$.

The dynamic structure factor in Eq.~(\ref{eq:DRP}) can also be
expressed as
\be
 S(\Delta {\bf K},\omega) = \frac{1}{2 \pi}
  \int e^{-i\omega t} \ \! I(\Delta{\bf K},t) \ \! dt ,
 \label{eq:DSF}
\ee
where
\be
 I(\Delta {\bf K},t) \equiv
  \langle e^{-i\Delta {\bf K} \cdot
   [{\bf R}(t) - {\bf R}(0)] } \rangle
  = \langle e^{-i\Delta K
    \int_0^t { v}_{\Delta{\bf K}} (t') \ \! dt'} \rangle
 \label{eq:IntSF}
\ee
is the so-called intermediate scattering function.
In this function, the brackets denote the ensemble average over
trajectories ${\bf R}(t)$ running on the surface, and
${v}_{\Delta{\bf K}}$ is the velocity of the adparticle
projected onto the direction of the parallel momentum transfer,
$\Delta {\bf K}$ ($\Delta K \equiv \| \Delta {\bf K} \|$).
Following the standard formulation used in neutron scattering theory,
this function can be split up into two contributions: $I_s$ ({\it self
function}), which corresponds to the average of trajectories of the
same adparticle at two different times, and $I_d$ ({\it distinct
function}), which corresponds to the average of trajectories of two
different adparticles at different times.
The Fourier transforms corresponding to $I$ and $I_s$ give
what are called the coherent and incoherent scattering laws,
$S(\Delta {\bf K},\omega)$ and $S_s (\Delta {\bf K},\omega)$,
respectively.

In QHAS experiments, and with interacting adsorbates, coherent
scattering is always obtained.
The corresponding theoretical interpretation of this type of
scattering is usually carried out in terms of Vineyard's
convolution approximation,\cite{vineyard} where the distinct pair
correlation function is expressed as a convolution of the self pair
correlation function.
This approximation is known to fail at small distances, where the
surface lattice becomes important.
Whereas in  neutron scattering many attempts to improve the convolution
approximation have been developed, within the QHAS context very little
effort has been devoted to achieve this goal.
At finite coverages, one usually distinguishes between two diffusion
coefficients:\cite{gomer} the tracer diffusion constant ($D_t$) and
the collective diffusion constant ($D_c$).
$D_t$ refers to the self-diffusion process and focuses on the motion
of a single adsorbate.
On the contrary, $D_c$ is related to the collective motion of all
adsorbates which is governed by Fick's law.
In either case, a Kubo-Green formula relates $D_t$ or $D_c$ with the
velocity autocorrelation function of a single adsorbate or with the
corresponding for the velocity of the center of mass, respectively.

In the ISA approximation, the distinction between self and distinct
time-dependent functions does not apply, and Eqs.~(\ref{eq:DRP})
and (\ref{eq:DSF}) still hold.
If the so-called Gaussian approximation\cite{gomer} is invoked, some
analytical treatment is possible.
This theoretical approach leads to a very simple and easy manner to
interpret the underlying dynamics issued from the numerical Langevin
simulations.
Under this approximation, the intermediate scattering function can be
expressed as a second-order cumulant expansion in $\Delta {\bf K}$ as
\be
 I(\Delta {\bf K},t) \approx
  e^{- \Delta  K^2 \int_0^t (t - t') \mathcal{C}(t') dt'} .
 \label{eq:IntSF2}
\ee
In Eq.~(\ref{eq:IntSF2}), $\mathcal{C}(t)$ is the velocity
autocorrelation function, defined as
\be
 \mathcal{C}(\tau) \equiv
 \langle v(0) v(\tau) \rangle
 \label{vcorr1}
\ee
along $\Delta {\bf K}$. The Gaussian approximation
is exact when the velocity correlations at more than two
different times are negligible, thus allowing to replace the average
acting over the exponential function by an average acting over its
argument.

The motion of an adsorbate under the action of a bath consisting
of other adsorbates on a static two-dimensional general surface
potential can be well described, in the Markovian
approximation,\cite{ruth1} by the standard Langevin equation
\begin{equation}
 \ddot{\bf R}(t) = - \eta \dot{\bf R}(t) + F[{\bf R} (t)] +
  \delta {\bf R}_S (t)+ \delta {\bf R}_G (t),
 \label{eq-lang1}
\end{equation}
where ${\bf R} = (x, y)$ represents the surface lattice points and
${\bf F} = - \nabla V$ is the deterministic force per mass unit,
with $V(x,y)$ being the surface interaction potential with
periods $a$ and $b$ along the $x$ and $y$ directions, respectively.
The substrate excitations lead to the random force per mass unit
${\bf R}_G(t)$ on the adatom which has features of a Gaussian white
noise.
The shot noise ${\bf R}_S (t)$ is given by a series of impacts, and we
have assumed sudden adsorbate-adsorbate collisions (i.e., strong but
elastic) and after-collision effects relax exponentially.
Moreover, the probability for collisions follows a Poisson
distribution.
No correlation has been assumed between both types of noise as well as
between the noise sources associated with each degree of freedom.

In Eq.~(\ref{eq-lang1}), $\eta = \gamma + \lambda$, where $\gamma$ is
the frictional damping coefficient resulting from the nonadiabatic
coupling to the electronic and vibrational excitations of the
substrate, and $\lambda$ is the number of collisions per time
unit or collisional friction.
By using the elementary kinetic theory of transport in
gases\cite{Mcquarrie} and the Chapman-Enskog theory for hard spheres,
a simple relation can be found\cite{ruth4} between the collisional
friction coefficient $\lambda$ and the coverage $\theta$ at a
temperature $T$,
\be
 \lambda = \frac{6 \rho \theta}{a^2} \ \! \sqrt{\frac{k_B T}{m}} ,
 \label{theta}
\ee
where $k_B$ is the Botzmann constant, $a$ is the unit cell length of
an assumed square surface lattice cell and $\rho$ is the effective
radius of an adparticle of mass $m$.

Finally, it is worth stressing the fact that, at long times, there
will not be free diffusion because the effects of the two stochastic
forces are dominant.
This is the {\it diffusive regime}, where mean square displacements
are linear with time,
\begin{equation}
 \langle x^2 \rangle (t) \sim \frac{2 k_B T}{m \eta} \ \! t = 2 D t ,
 \label{avvalues5}
\end{equation}
$D$ being the diffusion constant.
In general, we can express Eq.~(\ref{avvalues5}) as
\begin{equation}
 D = \lim_{t \to \infty} \frac{1}{4t}
  \langle |{\bf R} (t) - {\bf R} (0)|^2 \rangle ,
 \label{DD}
\end{equation}
which is known as {\it Einstein's law}, the value $D = k_B T/m\eta$ for
the diffusion coefficient being the {\it Einstein relation}.
Note from Eq.~(\ref{avvalues5}) that $D$ increases by lowering the
total friction $\eta$ and by increasing the surface temperature.


\section{Results and discussion}
\label{sec3}

The diffusion of Na atoms on the Cu(001) surface can be
considered as a prototype in QHAS since a lot of experimental and
theoretical work can be found in the literature.
The coverage $\theta_{Na} = 1$ corresponds to one Na atom per Cu(001)
surface atom\cite{toennies2} or, equivalently, $\sigma = 1.53 \times
10^{19}$~atom/cm$^2$; $a = 2.557$~\AA\ is the
unit cell length and $\rho = 2$~\AA\ has been used for the atomic
radius of Na. The surface friction we have considered
in our simulations is taken from Ref.~\onlinecite{toennies1} to be
$\gamma = 0.1 \ \! \omega_0 = 2.2049\times 10^{-5}$, where
$\omega_0$ is the harmonic frequency associated with the periodic
surface potential. Frequencies and times are here given in atomic
units.
We are going to analyze four coverages for two surface
temperatures: 200 and 300~K.
On the one hand, $\theta_{\rm Na} = 0.028$ and 0.064, where the single
adsorbate regime holds; on the other hand, $\theta_{\rm Na} = 0.106$
and 0.18, where the interaction between adsorbates cannot be ignored.
These coverage values correspond to $\lambda = 3.34 \times 10^{-6}$
($\eta = 2.53 \times 10^{-5}$), $\lambda = 7.64 \times 10^{-6}$
($\eta = 2.98\times 10^{-5}$), $\lambda = 1.26 \times 10^{-6}$
($\eta = 3.46\times 10^{-5}$) and $\lambda = 2.15 \times 10^{-5}$
($\eta = 4.68 \times 10^{-5}$), respectively.

For the adsorbate-substrate interaction potential $V$, in this work we
have considered the non-separable model proposed by Ellis {\it et
al.},\cite{toennies1} whose form is
\be
 V(x,y) = V_0(x,y) + V_1(x,y) + V_2(x,y) .
 \label{eq:PotTot}
\ee
In this expression, the first term is a simple separable cosine
potential,
\be
 V_0(x,y)=V_0 \, \left[2-\cos(2\pi x/a) - \cos(2\pi y/a) \right] ,
 \label{eq:Pot0}
\ee
with $a$ the lattice constant of the Cu(001) surface and
$V_0=41.4$~meV; the second term, which reads as
\be
 V_1(x,y) = - A \sum_{m,n}
  e^{-b \{[x/a-(m + 1/2)]^2 + [y/a-(n + 1/2)]^2\}} ,
 \label{eq:Pot1}
\ee
with $A=2V_0$ and $b=11.8$, is added to produce a lowering of the
potential barrier at on-top sites according to the observations;
finally, the third term is also a nonseparable part which serves to
alter the curvature near the minima and vary the difference between
the potential at the minima and the bridge positions,
\begin{eqnarray}
 V_2(x,y)& = & \pi^2 C V_0 \sum_{m,n}[(x/a-m)^2+(y/a-n)^2]
  \nonumber \\
  & & \times \exp[-(x/a-m)^2-(y/a-n)^2] ,
\label{eq:Pot2}
\end{eqnarray}
with $C=-0.2$. Two scattering azimuths, $[100]$ and $[1{\bar 1}0]$,
diagonal and parallel to the reciprocal space lattice in the first
Brillouin zone, respectively, are considered here.

\begin{figure}
 \includegraphics[width=7.5cm]{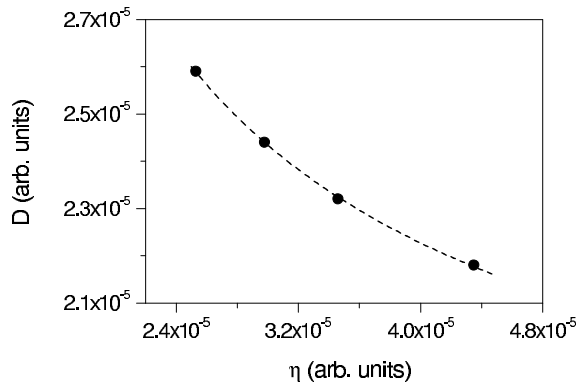}
 \caption{\label{fig1}
  Numerical (circles) and theoretical (dashed line) values for the
  diffusion coefficient $D$ as a function of the total friction $\eta$
  at $T = 200$~K along the azimuth $[100]$.}
\end{figure}

In the ISA approximation, Einstein's law, given by Eq.~(\ref{DD}), is
satisfied and, therefore, the diffusion constant $D$ decreases with the
coverage.
As said above, the origin of the friction arises from the collisions
among adsorbates and the influence of the substrate phonons.
This behavior is clearly observed in Fig.~\ref{fig1}, where two types
of calculations of the diffusion constant are plotted as a function
of the total friction (and, therefore, of the coverage, since the
surface friction coefficient is kept fixed in all our calculations)
and for a surface temperature of 200~K along the azimuth $[100]$.
The full circles correspond to the values obtained from Eq.~(\ref{DD})
after performing the corresponding Langevin numerical simulation.
The dashed line represents Einstein's relation with a fitting constant
instead of the actual value $k_B T / m$; the value obtained for such a
fitting constant would correspond to an effective mass 1.007 times the
Na mass.
As seen, the agreement is fairly good.

\begin{figure}
 \includegraphics[width=7cm]{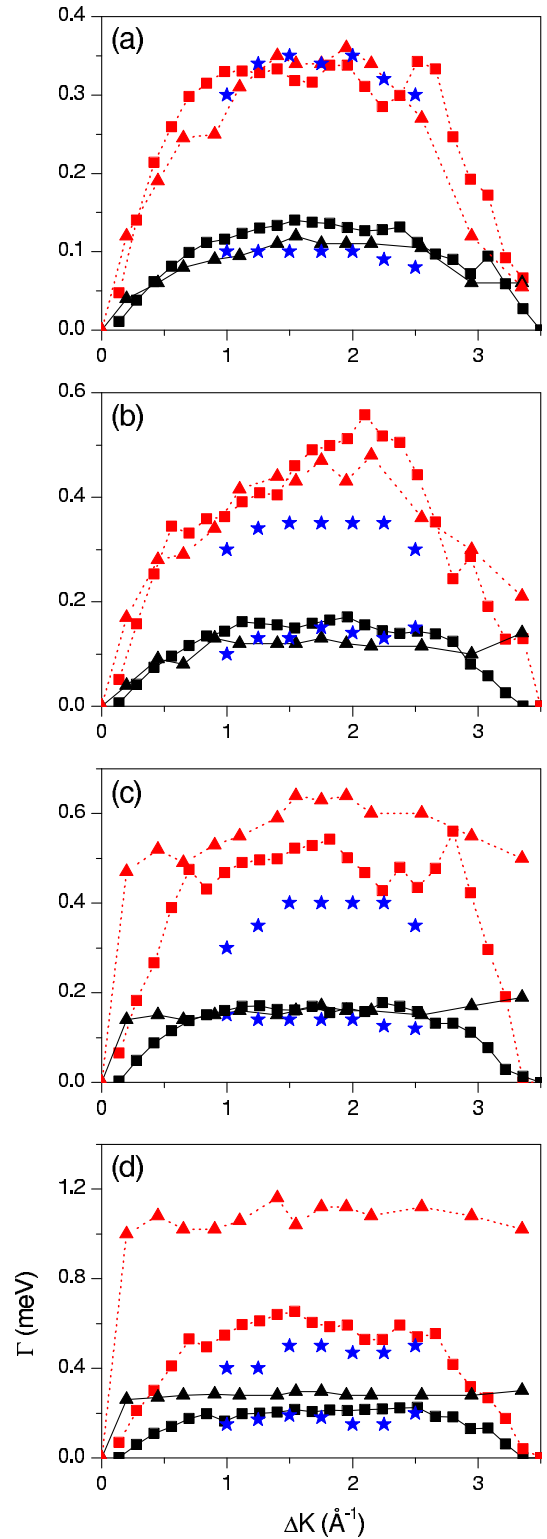}
 \caption{\label{fig2}
  (Color online.)
  Numerical (squares) and experimental (triangles) dependence of
  $\Gamma$ on $\Delta K$ at $T = 200$~K [black (solid) line] and
  $T = 300$~K [red (dotted) line] along the azimuth $[100]$.
  Different values of the coverage are considered: (a) $\theta=0.028$,
  (b) $\theta=0.064$, (c) $\theta=0.106$, and (d) $\theta=0.18$.
  To compare with, in each case a few LMD results (blue stars) are
  also shown.}
\end{figure}

\begin{figure}
 \includegraphics[width=7cm]{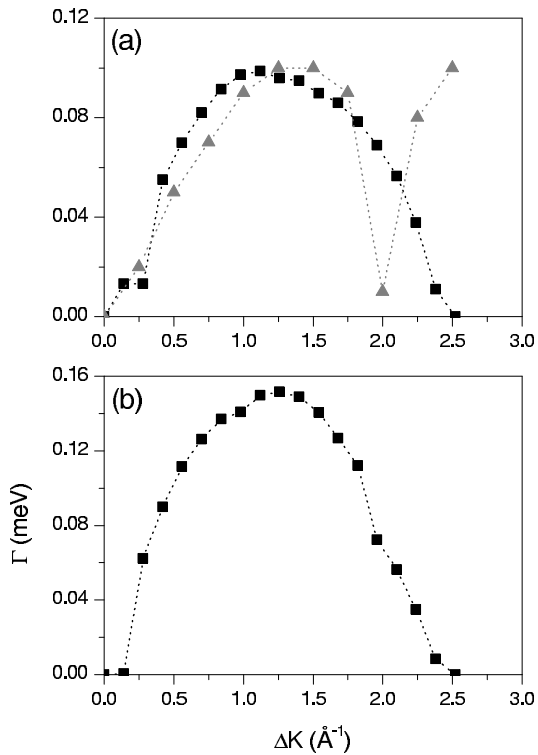}
 \caption{\label{fig3}
  Numerical (squares) and experimental (triangles) dependence
  of $\Gamma$ on $\Delta K$ at $T = 200$~K along the azimuth
  $[1{\bar 1}0]$.
  Two values of the coverage are considered: (a) $\theta=0.028$
  and (b) $\theta=0.18$.}
\end{figure}

The broadening observed in the {\it Q} peak with increasing coverage
can be attributed\cite{ruth3,ruth4,ruth5} to the respective fast decay
undergone by the intermediate scattering function.
The maximum value of the experimental peak widths is found to be
increased by a factor of 3 as the coverage is increased from 0.028
to 0.18.
In Figs.~\ref{fig2}(a)-(d), the results of our model for the Q-peak
width have been plotted and compared with the experimental ones
for $\theta=0.028$, $0.064$, $0.106$ and $0.18$ at $T= 200$~K [black
squares (solid lines)] and $T = 300$~K [red squares (dotted lines)]
together with the experimental ones (black and red triangles,
respectively) along the azimuth $[100]$.
To compare with, in each case a few LMD results are also shown (blue
stars).
As can be seen, LMD simulations reproduce the general trend,
but predict a smaller increase in the broadening of the {\it Q}
peak.\cite{toennies2}
However, the agreement of the ISA results with the experiment is fairly
good up to $\theta=0.106$.
This broadening has to be understood in terms of what we have termed
as {\it bound trajectories},\cite{ruth3} associated with the motion of
the adsorbates inside the potential wells: when adsorbates display
this type of motion, the complex exponentials in Eq.~(\ref{eq:IntSF})
keep the coherence necessary for the intermediate scattering function
to decay much slower than when the adsorbates are moving on the surface
({\it running trajectories}).
This effect is directly related to the increment of $\eta$, which
is equivalent to consider an increment of the collisional friction
$\lambda$ (which increases with $\theta$), since we have chosen
$\gamma$ as a constant substrate friction in all our numerical
simulations.
The same trend could be observed in the lineshapes associated with
the {\it T}-mode peaks.\cite{ruth3,ruth4,ruth5}

As mentioned above, the agreement of our model with the experiment is
at about $\theta=0.106$.
It is clear then that this value should be regarded as a limit for our
model.
Note that with high coverages, the adsorbate-adsorbate interaction
starts to play a more relevant role in the diffusion dynamics.
The average motion of the adsorbate becomes slower and it feels the
force exerted by its neighbors for longer times, which could end up
in a sort of collective motion involving a number of adsorbates.
This would give rise to a breakdown of the Markovian approximation
and, therefore, taking into account memory effects.

Broadening with the coverage can also be observed along the
$[1{\bar 1}0]$ azimuth.
As far as we know, no LMD simulations have been performed in this
direction.
Our numerical results (black squares) at two coverages ($\theta = 0.028$
and 0.18) and the experimental data (triangles) at $\theta = 0.028$ are
displayed in Figs.~\ref{fig3}(a) and \ref{fig3}(b).
In particular, the results for the highest value of the coverage are
considered to be predictive.


\section{Conclusions}
\label{sec4}

A full but simple stochastic description of the lineshapes for
activated surface diffusion and low-frequency vibrational motion for
adsorbate/substrate systems with increasing coverages has been
presented and discussed.
A simple Langevin equation within the ISA approximation is used to
calculate the adparticle dynamics, where a Gaussian white noise
accounting for the effect of the adsorbate-substrate friction and a
white shot noise simulating the friction due to the collisions among
adsorbates, with no correlation between them, are considered to
describe microscopic interactions.
In our opinion, the present numerical results for the broadening of
the {\it Q} peak, which are obtained with no fitting parameter, are
in better agreement than previous LMD simulations, mainly due to the
role of the collisional friction introduced in our stochastic model.
We are well aware that the barrier diffusion should be modified with
the coverage.\cite{toennies2}
However, our goal here was only to study the collisional friction
effects.
Obviously, this type of studies cannot replace further investigations
at microscopic level, and calculations from first principles of this
dynamics need to be done.
The complementary view we propose in this work should remain valid
because the lineshapes of the diffusion and low-vibrational frequency
motion analyzed at very long times, where the number of collisions is
relatively large (or where the statistical limit is applicable), seem
to wipe out any trace of microscopic interactions.


\section*{Acknowledgments}

This work was supported in part by DGCYT (Spain) under Project
FIS2007-62006.
R.M.-C.\ would like to thank the University of Bochum for support
from the Deutsche Forschungsgemeinschaft, SFB 558.
A.S.S. would like to thank the Spanish Ministry of Education and
Science for a ``Juan de la Cierva'' contract.



\begin{thebibliography}{99}
\eprint{}
\bibitem{salva1}
 S. Miret-Art\'es and E. Pollak, J. Phys.: Condens. Matter {\bf 17},
 S4133 (2005).

\bibitem{toennies1}
 A.P. Graham, F. Hofmann, J.P. Toennies, L.Y. Chen, and S.C. Ying,
 Phys. Rev. B {\bf 56}, 10567 (1997).

\bibitem{toennies2}
 J. Ellis, A.P. Graham, F. Hofmann, and J.P. Toennies,
 Phys. Rev. B {\bf 63}, 195408 (2001).

\bibitem{josele1}
 J.L. Vega, R. Guantes, and S. Miret-Art\'es,
 J. Phys.: Condens. Matter {\bf 14}, 6193 (2002);
 {\bf 16}, S2879 (2004).

\bibitem{josele2}
 J.L. Vega, R. Guantes, and S. Miret-Art\'es,
 Phys. Chem. Chem. Phys. {\bf 4}, 4985 (2002);
 R. Guantes, J. L. Vega, S. Miret-Art\'es, and E. Pollak,
 J. Chem. Phys. {\bf 119}, 2780 (2003).

\bibitem{josele3}
 R. Guantes, J. L. Vega, S. Miret-Art\'es, and E. Pollak,
 J. Chem. Phys. {\bf 120}, 10768 (2004);
 J. L. Vega, R. Guantes, S. Miret-Art\'es, and D. A. Micha,
 {\it ibid}. {\bf 121}, 8580 (2004).

\bibitem{ruth1}
 R. Mart\'{\i}nez-Casado, J.L. Vega, A.S. Sanz, and S. Miret-Art\'es,
 J. Phys.: Condens. Matter {\bf 19}, 176006 (2007).

\bibitem{ruth2}
 R. Mart\'{\i}nez-Casado, J.L. Vega, A.S. Sanz, and S. Miret-Art\'es,
 J. Chem. Phys. {\bf 126}, 194711 (2007).

\bibitem{allison}
 G. Alexandrowicz, A.P. Jardine, H. Hedgeland, W. Allison,
 and J. Ellis, Phys. Rev. Lett. {\bf 97}, 156103 (2006).

\bibitem{ruth3}
 R. Mart\'{\i}nez-Casado, J.L. Vega, A.S. Sanz, and S. Miret-Art\'es,
 Phys. Rev. Lett. {\bf 98} 216102 (2007).

\bibitem{ruth4}
 R. Mart\'{\i}nez-Casado, J.L. Vega, A.S. Sanz, and S.Miret-Art\'es,
 Phys. Rev. E {\bf 75}, 051128 (2007).

\bibitem{ruth5}
 R. Mart\'{\i}nez-Casado, J.L. Vega, A.S. Sanz, and S. Miret-Art\'es,
 J. Phys.: Condens. Matter {\bf 19}, 305002 (2007).

\bibitem{Mcquarrie}
 D.A. McQuarrie, {\it Statistical Mechanics}
 (Harper and Row, New York, 1976).

\bibitem{vanHove}
 L. van Hove, Phys. Rev. {\bf 95}, 249 (1954).

\bibitem{vineyard}
 G.H. Vineyard, Phys. Rev. {\bf 110}, 999 (1958).

\bibitem{gomer}
 R. Gomer, Rep. Prog. Phys. {\bf 53}, 917 (1990).

\end{thebibliography}
\end{document}